\documentclass[twocolumn,preprintnumbers,prd,superscriptaddress,nofootinbib,floatfix,showpacs,showkeys]{revtex4-1}

\usepackage{amsmath}
\usepackage{amsfonts}
\usepackage{amssymb}
\usepackage{graphicx}
\usepackage{color}
\usepackage{hyperref}
\usepackage{booktabs}
\usepackage{cleveref}
\usepackage{braket}
\usepackage{mathtools}
\usepackage[rightcaption]{sidecap}
\usepackage{subfigure}
\usepackage{comment}

\definecolor{vividviolet}{rgb}{0.62, 0.0, 1.0}
\definecolor{amaranth}{rgb}{0.9, 0.17, 0.31}
\definecolor{palatinateblue}{rgb}{0.15, 0.23, 0.89}
\definecolor{brightpink}{rgb}{1.0, 0.0, 0.5}
\definecolor{cornflowerblue}{rgb}{0.39, 0.58, 0.93}
\definecolor{deepcarminepink}{rgb}{0.94, 0.19, 0.22}
\definecolor{radicalred}{rgb}{1.0, 0.21, 0.37}

\hypersetup{ linktoc=all,
    colorlinks, linkcolor={palatinateblue},
    citecolor={brightpink}, urlcolor={amaranth}
}

\newcommand{\be}{\begin{equation}}
\newcommand{\ee}{\end{equation}}
\newcommand{\bs}{\begin{split}}
\newcommand{\bea}{\begin{eqnarray}}
\newcommand{\eea}{\end{eqnarray}}

\newcommand{\bes}{\begin{subequations}}
\newcommand{\ees}{\end{subequations}}

\graphicspath{{Graphics/}}

\def\be{\begin{equation}}
\def\ee{\end{equation}}
\def\bea{\begin{eqnarray}}
\def\eea{\end{eqnarray}}

\def\nat{Nature}

\def\prd{Phys. Rev. D}
\def\mnras{MNRAS}
\def\aj{AJ}
\def\apj{ApJ}

\def\apjs{ApJ Suppl. Ser.}

\def\physrep{Phys. Rep.}
\def\jcap{JCAP}

\def\apss{Astrophysics and Space Science}

\begin{document}

\title{Unifying the dark sector through a single matter fluid with non-zero pressure}

\author{Peter K. S. Dunsby}
\email{peter.dunsby@uct.ac.za}
\affiliation{Department of Mathematics and Applied Mathematics, University of Cape Town, Rondebosch 7701, Cape Town, South Africa.}
\affiliation{Cosmology and Gravity Group (CGG),
University of Cape Town, Rondebosch 7701, Cape Town, South Africa.}
\affiliation{South African Astronomical Observatory, Observatory 7925, Cape Town, South Africa.}
\affiliation{Centre for Space Research, North-West University, Potchefstroom 2520, South Africa}

\author{Orlando~Luongo}
\email{orlando.luongo@unicam.it}
\affiliation{Universit\`a di Camerino, Via Madonna delle Carceri 9, 62032 Camerino, Italy.}
\affiliation{SUNY Polytechnic Institute, 13502 Utica, New York, USA.}
\affiliation{Istituto Nazionale di Fisica Nucleare, Sezione di Perugia, 06123, Perugia,  Italy.}
\affiliation{INAF - Osservatorio Astronomico di Brera, Milano, Italy.}
\affiliation{NNLOT, Al-Farabi Kazakh National University, Al-Farabi av. 71, 050040 Almaty, Kazakhstan.}

\author{Marco Muccino}
\email{marco.muccino@lnf.infn.it}
\affiliation{Universit\`a di Camerino, Via Madonna delle Carceri 9, 62032 Camerino, Italy.}
\affiliation{NNLOT, Al-Farabi Kazakh National University, Al-Farabi av. 71, 050040 Almaty, Kazakhstan.}

\begin{abstract}
We explore a generalised unified dark energy model that incorporates a non-minimal interaction between a tachyonic fluid and an additional scalar field. Specifically, we require that the second field possesses a vacuum energy, introducing an ineliminable offset due to a symmetry-breaking mechanism. After the transition (occurring as due to the symmetry-breaking mechanism of the second field), the corresponding equation of state (EoS) takes the form of a combination between a generalised Chaplygin gas (GCG) component and a cosmological constant contribution. We reinterpret this outcome by drawing parallels to the so-called Murnaghan EoS, widely-employed in the realm of solid-state physics to characterise fluids that, under external pressure, counteract the pressure's effect. We examine the dynamic behaviour of this model and highlight its key distinctions compared to the GCG model. We establish parameter bounds that clarifies the model's evolution across cosmic expansion history, showing that it, precisely, exhibits behaviour akin to a logotropic fluid that eventually converges to the $\Lambda$CDM model in the early universe, while behaving as a logotropic or Chaplygin gas at intermediate and late times respectively. We explain our findings from a thermodynamic perspective, and determine the small perturbations in the linear regime. At very early times, the growth factor flattens as expected while the main departures occur at late times, where the Murnagham EoS results in a more efficient growth of perturbations. We discuss this deviation in view of current observations and conclude that our model is a suitable alternative to the standard cosmological paradigm, introducing the concept of a matter-like field with non-zero pressure.
\end{abstract}

\maketitle

\section{Overview}

Assuming that on large scales, the geometry of the universe is described by a Robertson-Walker metric, observations have conclusively shown that the expansion of the universe is currently accelerating \cite{1998Natur.391...51P,1998AJ....116.1009R,1999ApJ...517..565P,2003ApJ...594....1T,2003Sci...299.1532B,2003ApJS..148....1B,2003ApJS..148..135H,2003ApJS..148..161K,2003ApJS..148..175S,2005ApJ...633..560E}.
This phenomenon is commonly attributed to the thermodynamic properties of an exotic fluid with an unknown negative EoS that yields an effective repulsive gravity \cite{2000IJMPD...9..373S,2006IJMPD..15.1753C,tsujikawa2011dark}.
Clearly, this cannot be associated with ordinary fluids, such as matter and radiation, where the pressure is zero  and cannot drive the universe to accelerate today.

Thus, the common belief is that there exists an additional exotic fluid which is responsible for the cosmic speed-up \cite{2006IJMPD..15.1753C,2009IJMPA..24.1545M,2012Ap&SS.342..155B}, whose nature is currently attributed to the cosmological constant, $\Lambda$ \cite{2003PhR...380..235P} or, more broadly to some sort of dark energy contribution \cite{2003RvMP...75..559P}. However, this scenario is absolutely not exhaustive yet and, in fact, the precise nature of this fluid remains very poorly understood \cite{capozziello2013cosmographic}. Several conjectures have been put forward to explain the origins of the acceleration and identify the constituents responsible for it \cite{2009PhRvD..80b3008L,2020EPJC...80..629P,2003PhRvD..67f3504B,2007IJGMM..04..209S,2017PhR...692....1N,2018FoPh...48...17L,2013JCAP...07..017B}.

As stated above, the cosmological constant appears as the principal candidate responsible for the cosmic speed-up, making the concordance model statistically favoured \cite{burgess2015cosmological} as it depends on one parameter only, namely the cosmic mass. However,  there are still conceptual and theoretical issues raised by observed cosmological tensions \cite{weinberg1989cosmological},
which are yet to be solved. Nevertheless, dark energy remains a plausible explanation for these issues \cite{gruber2014cosmographic}. Unfortunately, having an evolving EoS also provides non-equilibrium effects and, moreover, we do not have direct evidence for dark energy reconstruction.

Additionally, the nature of dark matter provides a further challenge for theoretical cosmology \cite{2021PrPNP.11903865A}. It is known to interact through gravity, but its properties beyond that, i.e., microphysics, scattering properties, structure, etc., are still not well-understood \cite{2018Natur.562...51B}. Several candidates have been proposed, such as ultralight fields, extremely massive particles, geometric contributions, extensions of Einstein's gravity and so forth \cite{2021PrPNP.11903865A,2019Univ....5..213P} albeit no direct evidence has been found so far \cite{2020Symm...12.1648P}. Consequently, the search for dark matter particles is ongoing, through experiments involving  direct and indirect detection. Discovering dark matter origin would represent one of the most significant breakthroughs in our understanding toward the universe and its constituents.

Even though dark energy and dark matter are often considered as separate and completely different components \cite{del2012three}, some theories propose a unified approach in which a single dark fluid can explain both phenomena\footnote{This idea is similar to the concept of extended and/or modified gravity scenarios \cite{buchdahl1970non}, where an additional fluid can be inferred from the field theory.} \cite{becca2007dynamics}.  Recently, unified models, also known as \emph{unified dark energy or dark matter models}, have been proposed to eliminate the degrees of freedom due to quantum fluctuations in the early universe \cite{2012IJMPD..2130002Y}. These models have been characterised by barotropic fluids or scalar fields \cite{avelino2008linear} and the main assumption is that the fluid is a single entity, rather than a sum of dark matter and dark energy, and its net pressure is negative enough to drive the universe's present-day acceleration \cite{bento1999compactification}. Indeed, unifying dark energy and dark matter has two significant advantages: first, it requires only a single component to explain both the observed accelerated expansion and structure formation; second, it enables us to treat dark matter and dark energy at the perturbation level in the same way \cite{hu1999structure}. The prototype of such models is the Chaplygin gas \cite{2006tmgm.meet..840G}. These models have however been severely criticised in the last decade \cite{2011PhLB..694..289F}. On the other hand, dark fluids \cite{2011PhRvD..84h9905A} and logotropic models \cite{2015EPJP..130..130C} are also possible alternative toward unifying dark energy and dark matter, however, again, a pure logotropic model cannot describe the universe's  dynamics as shown in Ref.~\cite{2021PhRvD.104b3520B}, while a dark fluid needs additional explanations, as shown in Ref. \cite{2014IJMPD..2350012L,2018PhRvD..98j3520L}.

Motivated by these ideas, we explore how to construct a single dark fluid with almost negligible pressure at early times, but negative pressure at late times. To do so, we assume a tachyonic field minimally-coupled to  a further scalar field carrying  vacuum energy. The underlying scenario is to incorporate vacuum energy under the form of a quantum field cosmological constant into a tachyonic field that provides similar behaviour than previous unified dark energy models, such as the Chaplygin gas. We obtain this way a Chaplygin-like gas that includes quantum fluctuations as due to a symmetry breaking mechanism associated with the field transporting vacuum energy. Consequently, we find that an effective dark matter field arises, reducing at late times to a genuine cosmological. Contrary to the standard $\Lambda$CDM paradigm, our model induces a net pressure at early times that influences structure formation. Hence, the coincidence problem appears to be resolved,  since dark energy evolves in time, i.e., the corresponding effective framework works as a generalisation of the Chaplygin gas with vacuum energy. Afterwards, we search for a physical interpretation of our fluid and demonstrate that our EoS is akin to the \emph{Murnaghan fluid} \cite{murnaghan1944compressibility}, where  a \emph{single matter fluid} with non-zero EoS shows dust and dark energy as the cosmological scale changes. In particular, this implies that matter is not described by a dust-like fluid, but it behaves differently on cosmic scales. The Murnaghan fluid, initially proposed in contexts of solid state physics,  is therefore applied to the universe at different scales. To this end, the fluid naturally invokes the existence of a negative EoS we first physically interpret it in view of the background dynamics and then investigate how it affects the clustering of structures at early times.

Different stages are thus investigated to get constraints on the free parameters of the model, showing a good compatibility with observations. To do so, we work out the most recent type Ia supernovae (SNe Ia), baryonic acoustic oscillation (BAO) and observational Hubble data (OHD) catalogs to perform  Markov chain Monte Carlo (MCMC) simulations based on Metropolis-Hastings algorithm, and compare our findings with the standard background $\Lambda$CDM model. Further, we explore the thermodynamics of the Murnaghan fluid highlighting the main differences with respect to the Chaplygin gas.  Specifically, we show that there are regions in which the model predicts a logotropic behavior, emphasising the consequences on the observable universe, while having regions in which the fluid acts as a dark fluid. In other words, we conclude that our scenario predicts a unified model that incorporates previous frameworks into a single fluid of matter with pressure, making it a serious candidate for an alternative description of dark energy on large scales.

The paper is organised as follows. In Sect.~\ref{sezione2}, we describe how to incorporate vacuum energy into a tachyonic field coupling it to a further scalar minimally-coupled with the first one. In such a way, we introduce the basic demands of the Murnaghan fluid that is better described in Sect.~\ref{sezione3}, emphasising the corresponding thermodynamics of the fluid itself and the limiting cases predicted by our approach. In Sect.~\ref{sezione4}, we work out our numerical analysis. In Sect.~\ref{sezione5}, in view of our numerical analysis, we propose a physical interpretation of our fluid, discussing in particular the thermodynamic consequences of our recipe. Hence, introducing the concept of \emph{matter with pressure}, we study the impact of such a scenario on linear perturbations, in Sect.~\ref{sezione6}. Finally, in Sect.~\ref{sezione7}, we present the conclusion and perspectives of our work.

\section{Minimally-coupled tachyonic field with vacuum energy}\label{sezione2}

The existence of matter with non-zero pressure has posed a longstanding challenge in modern cosmology. This concept has been extensively explored in the frameworks of \emph{unified dark energy} models, where dark matter is characterised by a non-zero pressure.  These models offer a way to unify these two elusive components of the universe, where the primary unknown ingredient becomes solely dark matter. Noteworthy, such models can be derived from fundamental representations that consider the Lagrangian of specific fields associated with the dark matter constituent. By incorporating these unified models, we can potentially gain a deeper understanding of the nature and properties of dark matter and its interplay with dark energy.

A prototype of these models is represented by the dark fluid. Here, one has \emph{one fluid only}, determined by an EoS that resembles the total EoS induced by the $\Lambda$CDM model, where  instead, at late times, two fluids are involved, i.e., dust plus the cosmological constant $\Lambda$. A possible definition of dark fluid is recovered from the \emph{quasi-quintessence fluid} \cite{2022CQGra..39s5014D}, induced by an energy-momentum tensor of the form
\begin{subequations}
\begin{align}
\rho&=K(\partial \phi)+U(\phi)\,,\\
P&=-U(\phi)
\end{align}
\end{subequations}
where $U>0$ is the potential induced by a scalar field, whose generalized kinetic energy $K(\partial \phi)$ is a function of the velocity $\partial \phi\equiv\partial^\mu\phi$ of the scalar field, $\phi$. The Lagrangian representation can be written by means of $\mathcal L=K-U+\lambda Y$, with $\lambda$ the Lagrange multiplier. For additional details see Ref.~\cite{2018PhRvD..98j3520L}.
The quasi-quintessence fluid is capable of solving the cosmological constant problem, as demonstrated in Ref.~\cite{2022CQGra..39s5014D} and so it appears a viable alternative to the standard background model \cite{2018PhRvD..98j3520L,Belfiglio:2023eqi,2023arXiv230704739B}.

Even though quite relevant, the dark fluid is not the unique example of unified models of dark energy. Among all the other possibilities, some relevant approaches are the well-established Chaplygin gas or its generalisations  \cite{2002PhRvD..66d3507B}.
These models have been extensively investigated at late and early times, providing a fundamental representation in terms of tachyonic fields \cite{2011GrCo...17..259F}.

Motivated by such considerations, an interesting extension of the above models may also consider the case in which a tachyonic fluid transports vacuum energy, in analogy to the cosmological constant. To do so, we demand that a non-minimally coupled tachyonic field incorporates vacuum energy by means of a further field transporting it.

Thus, we conjecture the existence of a Lagrangian able to feature both a GCG model and a cosmological constant and consider
\begin{equation}
\mathcal L_{\rm s}=b(\phi)f(X)\,,
\end{equation}
where $X\equiv \partial_\sigma\phi\partial^\sigma\phi$ is the kinetic term of the field $\phi$ and $b(\phi)$ and $f(X)$ are analytical functions. Given the action $\hat S = \int \mathcal L_{\rm s} \sqrt{-g}dx^4$, the energy-momentum tensor reads
\begin{equation}
T_{\mu\nu}=-\frac{2}{\sqrt{-g}}\frac{\delta \hat S}{\delta g^{\mu\nu}}=-2\dfrac{\delta\mathcal L}{\delta g^{\mu\nu}}+g_{\mu\nu}\mathcal L\,,
\end{equation}
and the four-velocity $u_\mu =\partial_\mu\phi/\sqrt{X}$, the pressure and the density of this fluid become
\begin{subequations}
\begin{align}
         \rho_{\rm s} &= b(\phi)f(X)-2X b(\phi) f_X(X)\,,\\
        P_{\rm s} & = -b(\phi)f(X)\,,
\end{align}
\end{subequations}
where $f_X$ labels the derivative of $f$ with respect to the kinetic term.

It turns out that, in this framework, a constant term in the EoS cannot be recovered either assuming algebraic equations for $b(\phi)$ and $f(X)$, or by assuming $b=b(\phi,\nabla \phi)$ and $f=f(X,\Box X)$.

Nevertheless, assuming we modify $P_{\rm s}$ to include a constant term, or more generally a $l(\phi)$ contribution, and get the GCG simultaneously,  would lead to a modification in the Lagrangian of the form $\mathcal L_{\rm s}\rightarrow\mathcal L_{\rm s}-l(\phi)$ that however disagrees with the requirement to shift $P_{\rm s}\rightarrow P_{\rm s}+l(\phi)$. Consequently, there is no chance to include into a GCG a pure constant contribution working out with one single dynamical field, identified by $\phi$ and $X$.
\subsection{Double field approach}
As a byproduct of our conjecture, we justify the need of introducing a double-field, non-coupled Lagrangian of the form
\begin{equation}
\mathcal L = b(\phi)f(X)+Y-V^{{\rm eff}}(\psi)\,,
\end{equation}
where $\psi$ is the second field,  $Y=\partial_\sigma\psi\partial^\sigma\psi$ is the corresponding kinetic term, and $V^{{\rm eff}}=V_0+V_1(\psi^2-v^2)^2$ is assumed to be a symmetry-breaking potential, with $V_0$ and $V_1$ being positive-definite, that select, among all possible configurations acquired by $\psi$, the minimum $v$. In addition, to reproduce the GCG-like part of the EoS, we consider \cite{2011GrCo...17..259F}
\begin{equation}
f(X)=\left(1-X^{\frac{1+\alpha}{2\alpha}}\right)^{\frac{1+\alpha}{\alpha}}\,.
\end{equation}
By varying the corresponding action with respect to the metric tensor $g_{\mu\nu}$, we obtain the energy-momentum tensor
\begin{equation}
T_{\mu\nu} = \frac{b(\phi)X^\frac{1 - \alpha}{2\alpha}}{\left(1 - X^\frac{1 + \alpha}{2\alpha}\right)^\frac{1}{1 + \alpha}}\partial_\mu \phi\partial_\nu \phi -2 \partial_\mu\psi \partial_\nu\psi + \mathcal L\,g_{\mu\nu}\,.
\end{equation}
A direct comparison with the perfect fluid energy-momentum tensor provides
\begin{subequations}
\label{fin}
\begin{align}
\rho &= \frac{b(\phi)}{\left(1 - X^\frac{1 + \alpha}{2\alpha}\right)^\frac{1}{1 + \alpha}}-2Y-V_0-V_1(\psi^2-v^2)^2,\\
P &= - b(\phi)\left(1 - X^\frac{1 + \alpha}{2\alpha}\right)^\frac{\alpha}{1 + \alpha}+V_0+V_1(\psi^2-v^2)^2,
\end{align}
\end{subequations}
where the four-velocities for the fields $\phi$ and $\psi$ are given by $u_\mu = \partial_\mu \phi/\sqrt{X}$ and $v_\mu = \partial_\mu \psi/\sqrt{Y}$, respectively.

Hence, once the symmetry is broken, namely when  $\phi\rightarrow v$ and $Y\rightarrow0$, by introducing a shift of the density $\rho\rightarrow\rho-V_0$, Eqs.~\eqref{fin} combine into and reduce to the EoS
\begin{equation}
\label{fin2}
P = - \frac{b(\phi)^{1+\alpha}}{\rho^\alpha}+V_0\,.
\end{equation}
This, however, has been achieved by working out a shift of the  density that is valid only \emph{after the transition induced by the symmetry-breaking potential}.
In such a case we are able to freeze-out the kinetic term of $\psi$, that in general differs from $X$.

The above finding is manifestly a GCG with an additional constant term, reinterpreted as a cosmological constant, predicted by vacuum energy. Naively, because of the presence of the offset $V_0$, the model can exhibit regimes in which dark energy arises naturally, leading however to the existence of a matter fluid-like that provides pressure at all stages of the universe evolution.

In the next section, we discuss the physics of this fluid from a macroscopic perspective.

\section{Macroscopic interpretation of tachyonic fluid with bare cosmological constant term: The thermodynamic acceleraton}\label{sezione3}

Bearing Eq.~\eqref{fin2} in mind, we reinterpret it in terms of the so-called \emph{Murnaghan EoS}, that in standard thermodynamics is constructed with an additive constant contributing to a GCG and resembling the aforementioned approach that uses fields.

Specifically, in solid state physics, the Murnaghan EoS establishes a relationship between the volume $V$ and the pressure $P$ of a given physical system that features the behaviour of matter under high pressure.
Its thermodynamics reflects that experimentally the more a solid is compressed, the more difficult is to compress it further.
This process is related to characteristics such as Poisson coefficient, compressibility, etc. \cite{2015GReGr..47...63A}, and is extremely similar to what happens in the context of the Anton-Schmidt fluid \cite{2018PDU....20....1C}, recently characterised as an extensions of logotropic fluids \cite{Benaoum:2023ene}.

Thus, the basic assumption behind the Murnaghan EoS is that the bulk modulus of the incompressibility  $K=-V(\partial P/\partial V)_T$ at constant temperature $T$ is a linear function of pressure given by $K=K_0+K_0^\prime P$. This provides a pressure defined as
\begin{equation}\label{P}
    P = \frac{K_0}{K_0^\prime}\left[\left(\frac{V}{V_0}\right)^{-K_0^\prime}-1\right]\,,
\end{equation}
where $K^\prime$ is the first derivative of the bulk modulus with respect to $P$ and the subscript ``$0$'' labels the values of each quantity describing the system taken when $P=0$.

A straightforward application of the above EoS in describing the dynamics of the universe can be obtained by considering $V\propto\rho^{-1}$, where $\rho$ is the density.
However, key differences exist between solids and the universe.
\begin{itemize}
\item[-] For the universe, the condition $P=0$ occurs at intermediate/early times, when the pressure-less matter dominates the dynamics, namely, at a normalisation density $\rho_\star$ or at volume $V_0\propto\rho_\star^{-1}$.
\item[-] The universe expands, therefore, the volume $V_0\propto\rho_\star^{-1}$ is smaller than the volume $V\propto\rho^{-1}$ at pressure on-set, implying that $\rho_\star>\rho$.
\item[-] By definition $K>0$, so at $P=0$ it has to be $K_0>0$. Similarly, at the on-set of the cosmic pressure $P<0$ causing the accelerated expansion of the universe, to keep $K>0$ requires $K_0^\prime<0$.
\end{itemize}
In view of the above considerations, we may perform the substitutions $K_0\rightarrow A_\star$ and $K_0^\prime\rightarrow -\alpha$, where $\alpha>0$ and $A_\star>0$ are constants, therefore, ending up with
\begin{equation}\label{P1}
    P=-\frac{A_\star}{\alpha}\left[\left(\frac{\rho_\star}{\rho}\right)^{\alpha}-1\right]\,.
\end{equation}
Eq.~\eqref{P1} describes a Chaplygin-like behavior $P\propto \rho^{-\alpha}$ with an additional constant term, as above noticed.

Remarkably, we can notice that the original scalar field scenario assumes two fluids, identified by $\psi$ and $\phi$. However, in the Murnaghan picture we focus on one fluid only that unifies dark energy and darm matter under the same standards. While in the Murnaghan model a further constant contribution arises naturally, in the field representation the constant contribution comes from vacuum energy induced by $\psi$. Nevertheless, since $\psi$ falls in its minimum, the corresponding dynamics for $\psi$ is frozen. This is  analogous to claim that we obtain only one dynamical field, namely $\phi$, in analogy to the Murnaghan macroscopic representation of having one fluid only.

Interestingly, the magnitude of the constant term, mimicking the cosmological constant, is already tuned and can be compatible with both observations and Weinberg's no go theorem, see e.g. \cite{Nagahama:2018rcc}. We will focus on these points later in the text, discussing the cosmological constant problem in view of our framework.

\subsection{Large-scale dynamics}
By solving the continuity equation of a dark fluid with the above pressure,
\begin{equation}
    \label{cont_eq}
    \rho^\prime + 3(P+\rho)=0\,,
\end{equation}
where the prime represents the derivative with respect to $\ln{a}$ and $a$ is the scale factor, in principle, it is possible to obtain the dark fluid density $\rho$.

This strategy, however, does not provide any analytical solutions. Thus, to explore the feasibility of our  EoS, it would be useful to constrain the model parameters $A$, $\alpha$ and $\rho_\star$ to obtain approximate but meaningful solutions.

In other words, we will approximate our EoS, Eq.~\eqref{P1}, for different epochs, demanding that from each epoch we will bound the free parameters of our model.
\subsection{Mimicking a $\Lambda$CDM-like solution}
The first behaviour is associated with reproducing the $\Lambda$CDM scenario from our model. To this aim, we utilise the recent constraints on the GCG from Ref.~\cite{2022EPJC...82..582Z}, where quasars X-ray and UV flux measurements, \textit{Pantheon} sample of SNe Ia, compact radio quasars from very-long baseline interferometry, and  BAO data have been used. Ref.~\cite{2022EPJC...82..582Z} provides $\alpha=0.03^{+0.17}_{-0.14}$. However, in Eq.~\eqref{P1}, $\alpha<0$ implies $P>0$ and $\alpha=0$ implies $|P|\rightarrow\infty$; hence, it might be $\alpha\gtrsim0$. Since the scale factor $a$ is related to the redshift $z$ via $a=(1+z)^{-1}$, if $z\approx0$ or $a_0\approx1$ we may approximate $\rho\approx\rho_c$ -- where $\rho_c=3H_0^2/(8\pi G)$ is the universe's critical density, $H(a_0)\equiv H_0$ is the Hubble constant and $G$ is the gravitational constant -- to get
\begin{subequations}
\label{apprLCDM}
\begin{align}
\label{appr1}
\rho(z)&\approx \rho_c \left(\Omega_{\rm DF}+ \frac{P}{\rho_c}\right) a^{-3} - P\,,\\
\label{appr2}
P&\approx -A_\star \ln\left(\frac{\rho_\star}{\rho_c}\right)<0\,,
\end{align}
\end{subequations}
where $\Omega_{\rm DF}\equiv\rho/\rho_c$ is the dark fluid density parameter.

Remarkably, the above result is degenerate with the $\Lambda$CDM paradigm, as we required.
In fact, including the contributions of pressure-less baryonic matter and radiation, we obtain
\begin{equation}
\label{Hz}
H(a)=H_0\sqrt{\Omega_{\rm M}a^{-3} + \Omega_{\rm R}a^{-4} + \Omega_\Lambda}\,,
\end{equation}
where $\Omega_{\rm B}$, $\Omega_{\rm CDM}$, $\Omega_{\rm R}$, and $\Omega_\Lambda$ are the baryonic matter, the cold dark matter, the radiation, and the cosmological constant $\Lambda$ density parameters, respectively, and their relations are: $\Omega_{\rm M}\equiv\Omega_{\rm B}+\Omega_{\rm CDM}$, $\Omega_{\rm CDM}\equiv\Omega_{\rm DF} - \Omega_\Lambda$, and $\Omega_\Lambda\equiv-P/\rho_c$, and the flat universe prior provides $\Omega_{\rm M} + \Omega_{\rm R} + \Omega_\Lambda\equiv1$.

\subsection{Mimicking a logotropic-like solution}

If $K_0$ is constant or very weakly dependent upon $P$, then $K_0^\prime\approx0$ and from $K_0=-V(\partial P/\partial V)_T$, one gets $P=-K_0\ln(V/V_0)$. Resorting the usual substitutions to adapt the result to cosmological purposes, we get
\begin{equation}
\label{appr3}
P \approx -A_\star \ln \left(\frac{\rho_\star}{\rho}\right)<0\,,
\end{equation}
which generally holds for any value of $\rho$ and not only when $\rho\approx\rho_c$.
This result is remarkable, because it closely resembles the Anton--Schmidt EoS \cite{ANTON1997449,2019PhRvD..99b3532C,2021PhRvD.104b3520B}
\begin{equation}
\label{ASEoS}
P = A_n \left(\frac{\rho}{\rho_\star}\right)^{-n} \ln \left(\frac{\rho}{\rho_\star}\right)\,,
\end{equation}
with the index $n=0$, related to the Gr\"uneisen parameter $\gamma_{\rm G}=-1/6$ \cite{Gruneisen1912}, and the imposition $A_n\rightarrow A_\star$.
Strictly speaking, Eq.~\eqref{ASEoS} with the constraint $n=0$ is referred to as logotropic model \cite{2015EPJP..130..130C,2021PhRvD.104b3520B}.

From the first law of thermodynamics for adiabatic systems and Eq.~\eqref{appr3}, the energy density $\epsilon$ can be expressed in terms of the total matter density $\rho_{\rm M}(a)$ as
\begin{align}
\nonumber
\epsilon(a) =&\,\rho_{\rm M}(a) + \rho_{\rm M}(a) \int \frac{P \left[\rho_{\rm M}^\prime(a)\right]}{{\rho^\prime_{\rm M}(a)}^2} d \rho_{\rm M}^\prime(a)\\
=&\,\rho_{\rm M}(a) - A_\star \left\{1 + \ln\left[\frac{\rho_{\rm M}(a)}{\rho_\star}\right]\right\}\,.
\label{eq2}
\end{align}
By defining in Eq.~\eqref{eq2} the quantities
\begin{subequations}
\label{apprGL}
\begin{align}
B &\equiv [\ln(\rho_\star/\rho_{\rm M})-1]^{-1}\,,\\
A_\star&\equiv B\rho_c\Omega_\Lambda\,,
\end{align}
\end{subequations}
and dividing by $\rho_c$, so that $\rho_{\rm M}(z)/\rho_c\equiv \Omega_{\rm M}a^{-3}$, we obtain
\begin{equation}
\label{Hz_AS_logo}
H(a)=H_0\sqrt{\Omega_{\rm M} a^{-3} + \Omega_{\rm R} a^{-4} +  \Omega_\Lambda\left(1+3B\ln a \right)}\,,
\end{equation}
where we  introduced the contribution of the radiation and, knowing that $\Omega_{\rm M}\equiv\Omega_{\rm B} + \Omega_{\rm CDM}$, the flat universe prior provides $\Omega_{\rm M} + \Omega_{\rm R} + \Omega_\Lambda\equiv1$.

In Ref.~\cite{2016PhLB..758...59C}, the density $\rho_\star$, is identified with the Planck density $\rho_{\rm P}=c^5/(\hslash G^2)$, where $c$ is the speed of light and $\hslash$ is the reduced Planck constant. We refer to this case as GL1 model.  In general, $\rho_\star$ may be kept as a model parameter through $B$. We refer to this second case as GL2 model.
\subsection{Mimicking a generalized Chaplygin gas-like solutions}
Another interesting approximation of the Murnaghan EoS, that follows from the assumption of a bulk modulus with $K_0\ll K_0^\prime P$, is given by
\begin{equation}
\label{chaplygin}
P \approx \frac{K_0}{K_0^\prime}\left(\frac{V}{V_0}\right)^{-K_0^\prime}\,.
\end{equation}
By substituting $K_0\rightarrow \alpha A_c \rho_\star$ and $K_0^\prime\rightarrow -\alpha$, we obtain the GCG EoS \cite{2002PhRvD..66d3507B}
\begin{equation}\label{Pch}
    P=-A_c\rho_\star\left(\frac{\rho_\star}{\rho}\right)^\alpha\,.
\end{equation}
From Eq.~\eqref{cont_eq} we get
\begin{equation}
\Omega_{\rm DF}(a)= \Omega_{\rm DF} \left[A_s +(1-A_s)a^{-3(1+\alpha)}\right]^{\frac{1}{1+\alpha}}\,,
\end{equation}
with $\Omega_{\rm DF}(a)\equiv\rho(a)/\rho_c$ and $\Omega_{\rm DF}\equiv\rho(a_0)/\rho_c$. From the choices made for the model parameters, we have that
\begin{subequations}
\label{apprGCG}
\begin{align}
A_s&\equiv A_c\left[\frac{\rho_\star}{\rho(a_0)}\right]^{1+\alpha}\,,\\
A_\star&\equiv \alpha A_c \rho_\star\,.
\end{align}
\end{subequations}
Also, $A_s$ can be written in terms of the effective total matter density $\Omega_{\rm M}$ and $\alpha$ as \cite{2022EPJC...82..582Z}
\begin{equation}
\label{exprAs}
A_s = 1 - \left(\frac{\Omega_{\rm M}-\Omega_{\rm B}}{1-\Omega_{\rm B}}\right)^{1+\alpha}\,.
\end{equation}
Including baryonic matter and radiation, we get
\begin{equation}
\label{Hzch}
H(a)=H_0\sqrt{\Omega_{\rm B}a^{-3} + \Omega_{\rm R}a^{-4} + \Omega_{\rm DF}(a)}\,,
\end{equation}
where the flat universe prior gives $\Omega_{\rm B} + \Omega_{\rm R} + \Omega_{\rm DF}\equiv1$.

To focus on which epochs are related to the above approximations, we now intend to constrain our models and fix the bounds over the free parameters. We will thus conclude on how to interpret the model throughout the evolution of the universe.

\section{Numerical constraints}\label{sezione4}

In the concordance model, the total energy density can be decomposed into pressure-less (baryonic and dark) matter and cosmological constant $\Lambda$.
When the Murnaghan EoS is taken into account, the above decomposition is less trivial because the solution of the continuity equation is not analytic and our dark matter component does not exhibit a vanishing pressure.
Among the above analyzed approximated solutions of the Murnaghan EoS, we have seen that in $\Lambda$CDM-like and logotropic solutions dark energy and dark matter components are ``disentangled", whereas in the GCG solution these components are intertwined.

In view of these considerations, the study of the above approximated solutions can provide constraints for the model parameters of the full numerical solution of the Murnaghan EoS. Therefore, we perform a set of MCMC analyses to fix the cosmological bounds over the different paradigms. We employ the standard low-redshift data surveys: OHD \citep{2022LRR....25....6M}, the {\it Pantheon} catalog of SNe Ia \cite{2018ApJ...859..101S}, and BAO \cite{2019JCAP...10..044C}, all referring to the cosmological redshift $z$, which is easier to measure.
The best set of model parameters is set by maximising the total log-likelihood function
\begin{equation}
 \ln{\mathcal{L}} = \ln{\mathcal{L}_{\rm O}} + \ln{\mathcal{L}_{\rm S}} + \ln{\mathcal{L}_{\rm B}}\,.
\end{equation}
Below, we define the contribution of each probe.
\subsection{Hubble rate data log-likelihood}
OHD are cosmology-independent estimates of the Hubble rate $H(z)=-(1+z)^{-1}\Delta z/\Delta t$, obtained through spectroscopic measurements of the age difference $\Delta t$ and redshift difference $\Delta z$ of couples of passively evolving galaxies that formed at the same time \cite{2002ApJ...573...37J}.
The corresponding log-likelihood function is then given by
\begin{equation}
\label{chisquared1}
 \ln{\mathcal{L}_{\rm O}} = -\frac{1}{2} \sum_{i=1}^{N_{\rm O}} \left\{\ln{(2\pi\sigma_{H_i}^2)} + \left[\frac{H_i-H(z_i)}{\sigma_{H_i}}\right]^2\right\},
\end{equation}
where $N_{\rm O}=32$ corresponds to the OHD data points \cite{2022LRR....25....6M}.
\subsection{SNe Ia log-likelihood}
The \emph{Pantheon} data set of $1048$ SNe Ia \cite{2018ApJ...859..101S} is equivalently described as a $E(z)^{-1}\equiv H_0/H(z)$ catalog, at $N_{\rm S}=6$ redshifts, which accurately reproduce the cosmological constraints of the whole SN Ia data set \cite{2018ApJ...853..126R}.
The corresponding SN log-likelihood function is given by
\begin{align}
\nonumber
\ln \mathcal{L}_{\rm S} = & -\frac{1}{2}\sum_{i=1}^{N_{\rm S}} \left[E_i^{-1} - E(z_i)^{-1} \right]^{\rm T} \mathbf{C}_{\rm S}^{-1}
\left[E_i^{-1} - E(z_i)^{-1} \right]\\
\label{loglikeSN}
& -\frac{1}{2}\sum_{i=1}^{N_{\rm S}} \ln \left(2 \pi |\det\mathbf{C}_{\rm S}| \right)\,,
\end{align}
where $E_k^{-1}$ are the measurements from SNe Ia and $\mathbf{C}_{\rm S}$ is the covariance matrix \cite{2018ApJ...853..126R}.

\begin{table*}
\centering
\setlength{\tabcolsep}{.5em}
\renewcommand{\arraystretch}{1.3}
\caption{Best-fit and derived parameters (with $1$--$\sigma$ errors), and statistical tests of $\Lambda$CDM, GL1, GL2, and GCG models. As the Murnagham fluid cannot be integrated analytically, we separetely fit each of the above approximations, namely the $\Lambda$CDM, GL1, GL2 and GCG scenarios, to constrain the free parameters of our overall model, based on the non-minimal coupling between two scalar fields.}
\begin{tabular}{lccccccccrcc}
\hline\hline
                                    &
\multicolumn{4}{c}{Best-fit parameters}&
                                    &
\multicolumn{2}{c}{Derived parameters}&
                                    &
\multicolumn{3}{c}{Statistical tests}      \\
\cline{2-5}\cline{7-8}\cline{10-12}
Model                               &
$h_0$                               &
$\Omega_{\rm M}$                    &
$B$                                 &
$\alpha$                            &
                                    &
$\log(\rho_\star/\rho_c)$           &
$\log(A_\star/\rho_c)$              &
                                    &
$-\ln \mathcal L$                   &
AIC (BIC)                           &
$\Delta$                           \\
\hline
$\Lambda$CDM                        &
$0.690^{+0.012}_{-0.011}$           &
$0.304^{+0.028}_{-0.024}$           &
--                                  &
--                                  &
                                    &
$122.76$                            &
$-2.608^{+0.019}_{-0.020}$          &
                                    &
$65.50$                             &
$135$ ($139$) & $0$ ($0$)           \\
GL1                                &
$0.691^{+0.011}_{-0.011}$           &
$0.305^{+0.027}_{-0.025}$           &
--                                  &
--                                  &
                                    &
$122.76$                            &
$-2.610^{+0.019}_{-0.020}$       &
                                    &
$65.50$                             &
$135$ ($139$) & $0$ ($0$)           \\
GL2                                &
$0.711^{+0.012}_{-0.013}$           &
$0.358^{+0.031}_{-0.029}$           &
$0.319^{+0.042}_{-0.041}$           &
--                                  &
                                    &
$1.35^{+0.09}_{-0.12}$              &
$-1.135^{+0.038}_{-0.042}$          &
                                    &
$74.98$                             &
$156$ ($162$) & $21$ ($23$)         \\
GCG                                 &
$0.691^{+0.014}_{-0.013}$           &
$0.307^{+0.035}_{-0.028}$           &
--                                  &
$0.03^{+0.22}_{-0.17}$              &
                                    &
$122.76$                            &
$-4.971^{+0.019}_{-0.020}$          &
                                    &
$65.50$                             &
$137$ ($143$) & $2$ ($4$)           \\
\hline
\end{tabular}
\label{tab:results}
\end{table*}
\begin{figure*}
{\includegraphics[width=0.32\hsize,clip]{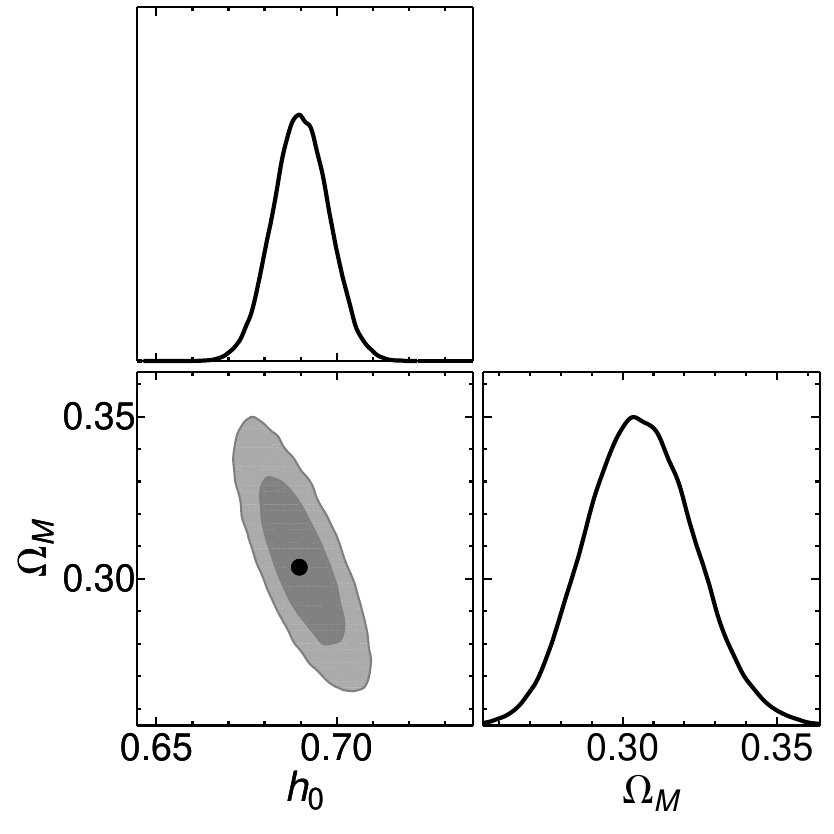}\hfill
\includegraphics[width=0.32\hsize,clip]{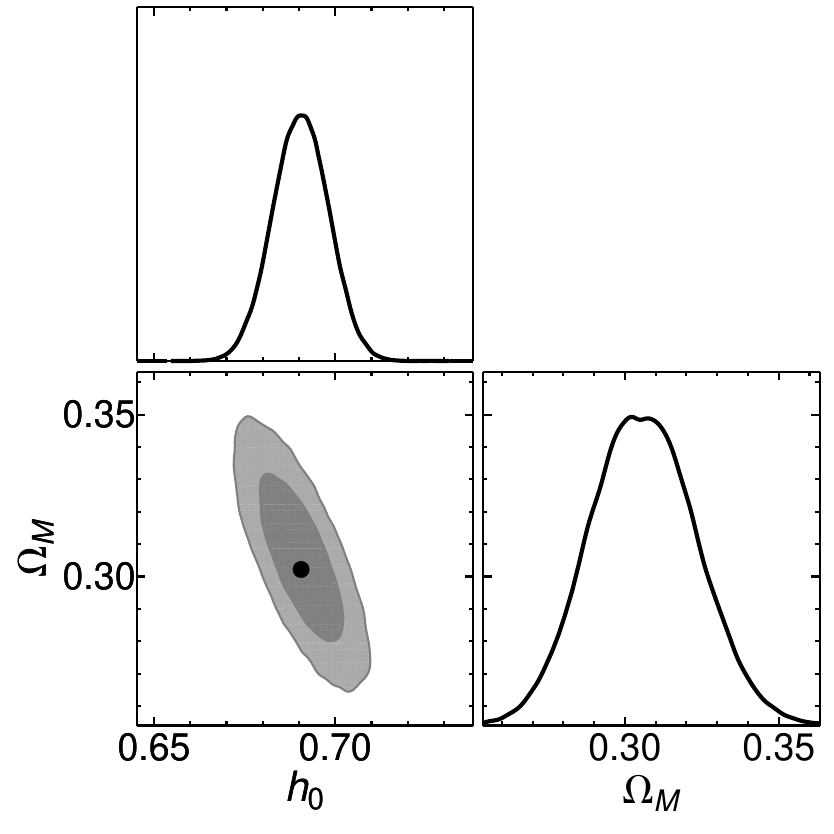}\hfill}

{\includegraphics[width=0.45\hsize,clip]{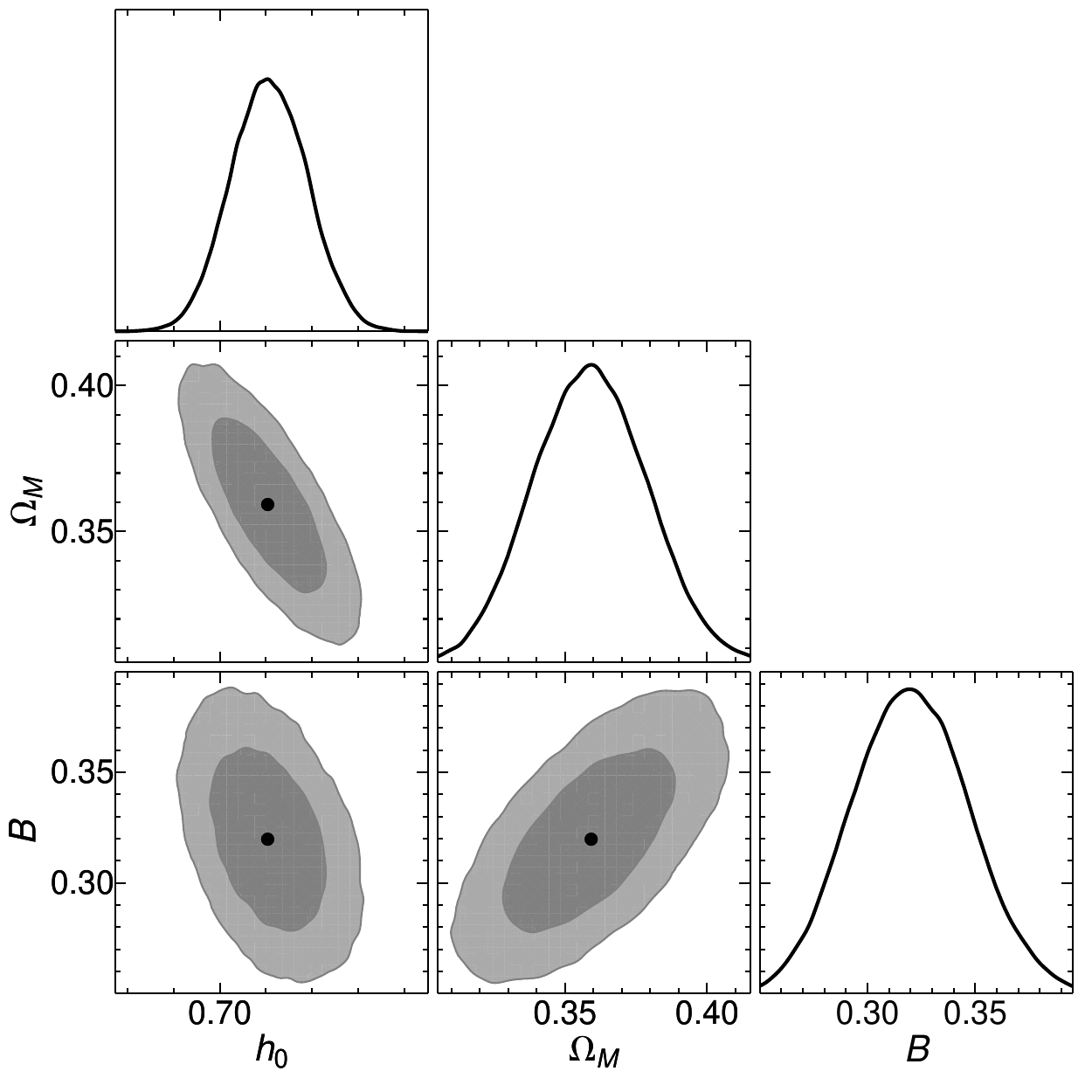}\hfill
\includegraphics[width=0.45\hsize,clip]{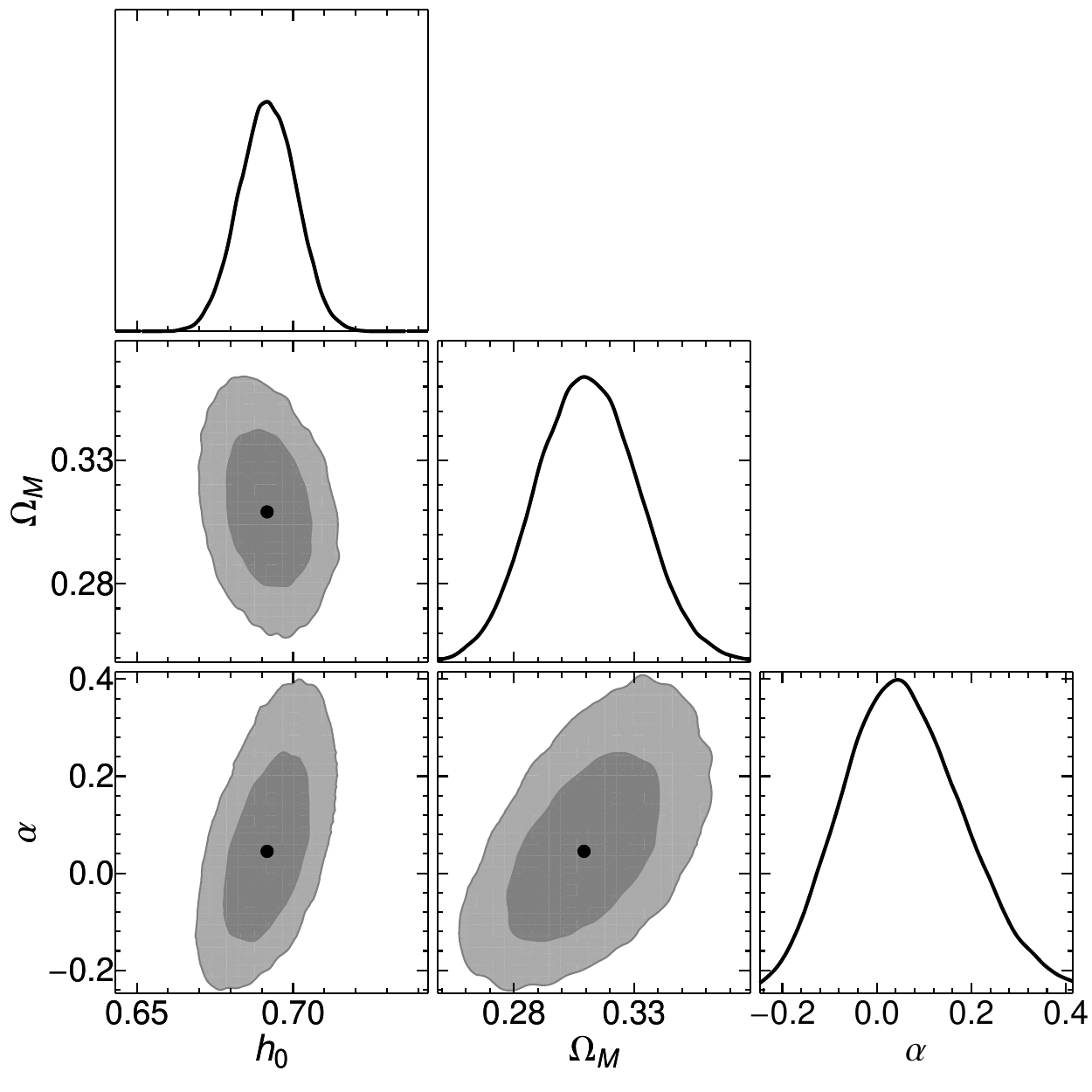}\hfill}
\caption{Contour plots and best-fit parameters (black dots) of $\Lambda$CDM (top left), GL1 (top right), GL2 (bottom left), and GCG (bottom right) models. Darker (lighter) areas mark $1$--$\sigma$ ($2$--$\sigma$) confidence regions.}
\label{fit:contours}
\end{figure*}
\subsection{BAO log-likelihood}
BAO are the observed peaks in the large scale structure correlation function and enable angular measurements in different redshift slabs that might be correlated or not.

For uncorrelated data points we define
\begin{equation}
\label{eq:DV}
\Theta(z) \equiv r_{\rm s} \left[\frac{H(z)}{cz}\right]^\frac{1}{3}\left[\frac{\left(1+z\right)}{d_{\rm L}(z)}\right]^\frac{2}{3}\,,
\end{equation}
where, for a flat universe, the luminosity distance is
\begin{equation}
\label{dlEz}
d_{\rm L}(z)=c(1+z)\int_0^z\dfrac{dz'}{H(z')}\,,
\end{equation}
and $r_{\rm s}$ is the comoving sound horizon at the baryon drag redshift, calibrated through the CMB data for a given cosmological model. Hereafter we fix $r_{\rm s}=147.09$~Mpc \cite{Planck2018}.
The corresponding log-likelihood is given by
\begin{equation}
\label{loglikebao}
\ln \mathcal{L}_{\rm B}^{\rm u} = -\frac{1}{2}\sum_{i=1}^{N_{\rm B}^{\rm u}}\left\{\ln (2 \pi\sigma_{\Theta_i}^2) + \left[\frac{\Theta_i - \Theta(z_i)}{\sigma_{\Theta_i}}\right]^2 \right\}\,,
\end{equation}
where the $N_{\rm B}^{\rm u}=8$ measurements are taken from Ref.~\cite{LM2020}.

For correlated data, we define
\begin{equation}
\label{eq:Az}
\Xi(z) \equiv \frac{H_0 r_{\rm s}}{cz}\frac{\sqrt{\Omega_{\rm m}}}{\Theta(z)}\,,
\end{equation}
and write the corresponding log-likelihood
\begin{align}
\nonumber
\ln \mathcal{L}_{\rm B}^{\rm c} = & -\frac{1}{2}\sum_{i=1}^{N_{\rm B}^{\rm c}} \left[\Xi_i - \Xi(z_i) \right]^{\rm T} \mathbf{C}_{\rm B}^{-1}
\left[\Xi_i - \Xi(z_i) \right]\\
\label{loglikeBAOc}
& -\frac{1}{2}\sum_{i=1}^{N_{\rm B}^{\rm c}} \ln \left(2 \pi |\det\mathbf{C}_{\rm B}| \right)\,,
\end{align}
where the $N_{\rm B}^{\rm c}=3$ measurements and the covariance matrix $\mathbf{C}_{\rm B}$ are taken from Ref.~\cite{2011MNRAS.418.1707B}.

Therefore, the full BAO log-likelihood is given by
\begin{equation}
\ln \mathcal{L}_{\rm B} = \ln \mathcal{L}_{\rm B}^{\rm u} + \ln \mathcal{L}_{\rm B}^{\rm c}\,.
\end{equation}
\subsection{Numerical results}
Tab.~\ref{tab:results} and Fig.~\ref{fit:contours} show, respectively, the contour plots and the best-fit parameters got from $\Lambda$CDM, GL1, GL2, and GCG models. We parametrised $H_0=100h_0$ and fixed $\Omega_{\rm R}=9.265\times10^{-5}$ and  $\Omega_{\rm B}=0.0493$ \cite{Planck2018}.

The statistical comparison of the above models is based on the \emph{Aikake Information Criterion} (AIC) and \emph{Bayesian Information Criterion} (BIC) \cite{2007MNRAS.377L..74L}, respectively,
\begin{subequations}
\begin{align}
{\rm AIC}&=-2\ln \mathcal L+2p\,,\\
{\rm BIC}&=-2\ln \mathcal L+p\ln N\,,
\end{align}
\end{subequations}
where $\ln \mathcal L$ is the maximum of the log-likelihood, $p$ is the number of parameters, and $N$ is the number of data points.
The model with the lowest values of the above criteria, labeled with AIC$_0$ and BIC$_0$, is referred to as the fiducial (best-suited) model.
The comparison between the proposed models and the fiducial one is performed by computing the differences $\Delta={\rm AIC}-{\rm AIC}_0$ or ${\rm BIC}-{\rm BIC}_0$, that provide evidence against the proposed model or, equivalently, in favour of the fiducial one, as follows
\begin{itemize}
    \item[-] $\Delta\in[0,\,3]$, weak evidence;
    \item[-] $\Delta\in (3,\,6]$, mild evidence;
    \item[-] $\Delta>6$, strong evidence.
\end{itemize}

Looking at the statistical test part of Tab.~\ref{tab:results}, we can conclude that $\Lambda$CDM and GL1 models are equally best-suited to fit the data, the GCG model is weakly/mildly disfavoured, whereas the GL2 model is strongly excluded.
\section{Physical interpretation of our double-field model}\label{sezione5}
From the results of the statistical analysis summarised Tab.~\ref{tab:results}, now we can provide handful constraints on the parameters ($\alpha$, $\rho_\star$, $A_\star$) of the Murnaghan EoS.
\begin{itemize}
    \item[-] {\bf $\Lambda$CDM}. For this limiting case, we assumed $\alpha\gtrsim0$. However, we are unable to constrain $\rho_\star$ and $A_\star$ because from Eqs.~\eqref{apprLCDM} we get only one condition
    \begin{equation}
         A_\star = \frac{3H_0^2}{8\pi G}\left[\ln\left(\frac{8\pi G\rho_\star}{3H_0^2}\right)\right]^{-1}\left(1-\Omega_{\rm M} - \Omega_{\rm R}\right)\,.
    \end{equation}
    In Tab.~\ref{tab:results}, to break the above-described degeneracy, we impose $\rho_\star\equiv\rho_{\rm P}$ and derive $A_\star$.
    \item[-] {\bf GL1} and {\bf GL2}. Also for these models we assumed $\alpha\gtrsim0$ but now, from Eqs.~\eqref{apprGL}, we are able to set conditions on both $\rho_\star$ and $A_\star$, i.e.,
    \begin{subequations}
    \begin{align}
    \label{rhostar}
    \rho_\star=&\frac{3H_0^2\Omega_{\rm M}}{8\pi G}\exp\left(1+\frac{1}{B}\right)\,,\\
        A_\star=&\frac{3H_0^2B}{8\pi G}\left(1-\Omega_{\rm M} - \Omega_{\rm R}\right)\,.
    \end{align}
    \end{subequations}
    For the GL1 model, the condition $\rho_\star\equiv\rho_{\rm P}$ holds, thus in Tab.~\ref{tab:results} we evaluated only $A_\star$. For the GL2 model, we evaluated both $\rho_\star$ and $A_\star$.
    \item[-] {\bf GCG}. In this case $\alpha$ is a model parameter and its best-fit value (see Tab.~\ref{tab:results}) agrees with the results got in Ref.~\cite{2022EPJC...82..582Z}.
    On the other hand, like in the $\Lambda$CDM case, from Eqs.~\eqref{apprGCG} we obtain only one condition
        \begin{equation}
        \label{Astar_GCG}
         A_\star = \alpha\frac{A_{\rm s}}{\rho_\star^\alpha}\left(\frac{3H_0^2}{8\pi G}\right)^{1+\alpha}\left(1-\Omega_{\rm B} - \Omega_{\rm R}\right)^{1+\alpha}\,,
    \end{equation}
    that does not allow us to disentangle $\rho_\star$ and $A_\star$ and the expression of $A_{\rm s}$ is given by Eq.~\eqref{exprAs}.
\end{itemize}

Now, we can draw some interesting considerations.
\begin{itemize}
    \item[-] At late times, when $a\approx1$, the Murnaghan EoS in Eq.~\eqref{P1} is approximated by the $\Lambda$CDM model for $\rho\approx\rho_{\rm c}$; the GL1 and GL2 models represent the limiting cases of Eq.~\eqref{P1} for $\rho\neq\rho_{\rm c}$.
    \item[-] Looking at Tab.~\ref{tab:results}, we notice that the GL1 is statistically identical to and degenerates with the $\Lambda$CDM model, whereas the model GL2 is strongly excluded. Therefore, since $\Lambda$CDM and GL1 models are limiting cases of Eq.~\eqref{P1}, we can safely deduce that the condition $\rho_\star\equiv\rho_{\rm P}$ holds also for the more general Murnaghan EoS.
    \item[-] At intermediate times $0\lesssim a\lesssim 1$, once the condition $\rho_\star\equiv\rho_{\rm P}$ is set-up, the term $(\rho_{\rm P}/\rho)^\alpha$ dominates and Eq.~\eqref{P1} can be approximated by the GCG model. Since this model is the only one able to constraint the parameter $\alpha$, we use the corresponding value listed in Tab.~\ref{tab:results} as a constraint for the Murnaghan EoS.
    \item[-] Putting together in Eq.~\eqref{Astar_GCG} the constraint $\rho_\star\equiv\rho_{\rm P}$ got from $\Lambda$CDM and GL1 models and the value of $\alpha$ got from the GCG model, we get the value of $A_\star$ listed in Tab.~\ref{tab:results} for the GCG case that represents the last constraint for the Murnaghan EoS.
    \item[-] Finally we notice that, at early times, when $a\approx0$ and $x=\rho_{\rm P}/\rho\approx1$, at the lowest order we have $x^\alpha - 1\approx\ln x^\alpha$, therefore, Eq.~\eqref{P1} becomes the EoS of the GL model in Eq.~\eqref{appr3}.
\end{itemize}

\begin{figure}
\includegraphics[width=\hsize,clip]{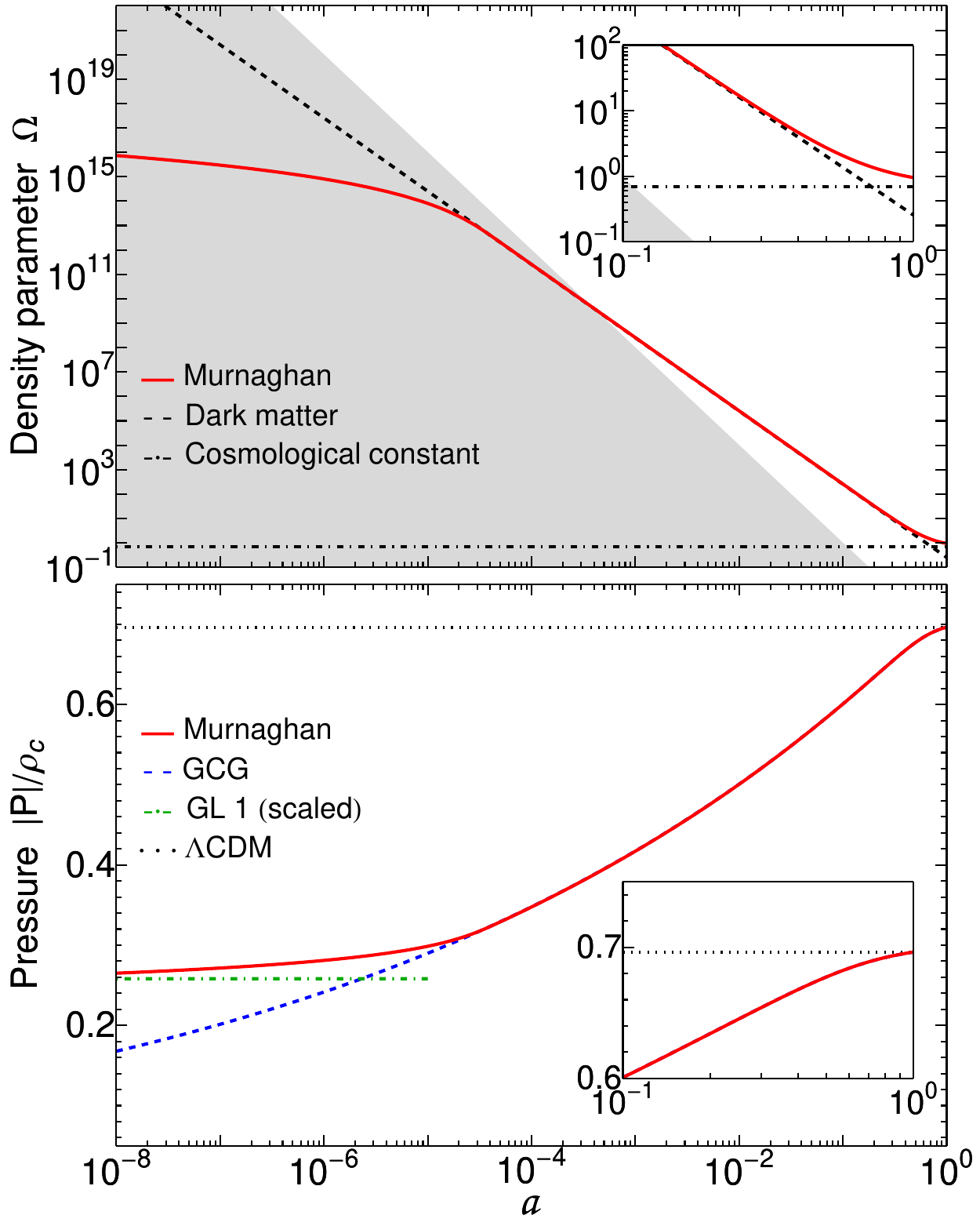}
\caption{\emph{Top panel}: numerical density from the Murnaghan EoS, compared with dark matter and dark energy behaviours of the $\Lambda$CDM case; the shaded area marks the radiation density behaviour. \emph{Bottom panel}: comparison among the pressures of Murnaghan, GCG, GL1 (rescaled to match the Murnaghan one at $a\rightarrow0$), and $\Lambda$CDM. The insets show the behaviours at $0.1\leq a\leq1$. The parameter values are taken from Tab.~\ref{tab:results}.}
\label{fig:EoS}
\end{figure}

In view of the above findings, viable constraints on the parameters of the Murnaghan EoS are:
\begin{subequations}
\label{consMurnpar}
\begin{align}
    \bar \alpha&=0.03^{+0.22}_{-0.17}\,,\\
    \bar\rho_\star&=\rho_{\rm P}\,,\\
    \bar A_\star&= \left(1.07^{+0.05}_{-0.05}\right)\times10^{-5}\rho_{\rm c}\,,
\end{align}
\end{subequations}
where we indicated with upper bars the parameters, pointing out that those limits are average quantities got from the fits, performed at different times.

The corresponding best-fit cosmological parameters, with associated errors, can be found from the numerical bounds obtained from $\Lambda$CDM, GL1 and GCG models, having:
\begin{subequations}
\label{consMurncos}
\begin{align}
    \bar h_0&=0.691^{+0.14}_{-0.13}\,,\\
    \bar\Omega_{\rm M}&=0.305^{+0.037}_{-0.026}\,.
\end{align}
\end{subequations}

In particular, bearing such intervals in mind, we notice that Fig.~\ref{fig:EoS}, top panel, displays the numerical density of the Murnaghan EoS for the parameters ($\bar\alpha$, $\bar\rho_\star$, $\bar A_\star$) listed in Eqs.~\eqref{consMurnpar} and the cosmological parameters ($\bar h_0$, $\bar\Omega_{\rm M}$) summarized in Eqs.~\eqref{consMurncos}. Being a dark fluid numerical solution, it is compared with the dark (matter and energy) sector behaviors obtained with the parameters ($h_0$, $\Omega_{\rm M}$) taken from the $\Lambda$CDM case in Tab.~\ref{tab:results}.

The comparison of the pressures of Murnaghan, GCG, GL1 (rescaled to match the Murnaghan one at $a\rightarrow0$), and $\Lambda$CDM are portrayed in the bottom panel of Fig.~\ref{fig:EoS}. The curves, obtained using the values of  Tab.~\ref{tab:results}, show explicitly that the full numerical solution behaves like
\begin{itemize}
    \item[-] $\Lambda$CDM and GL1 cases at $a\approx0$,
    \item[-] the GCG model at $10^{-5}\lesssim a\lesssim 1$, and
    \item[-] a scaled GL1 case at $a\lesssim10^{-5}$.
\end{itemize}
Hence, we find the physical interpretation below.
\begin{itemize}
    \item[-] At very early times, the model acts as logotropic fluid that reduces to the $\Lambda$CDM model. This implies that, in view of structure formation, we do not expect significant departures adopting our model. Phrasing it differently, a similar behaviour of our current standard cosmological model is expected even at the level of structure formation.
    \item[-] At early times, the model appears as a logotropic fluid, being a limiting case of the Anton-Schmidt equation of state. This is not in contrast with current knowledge about this epoch, albeit possible comparisons adopting high-distance indicators, such as gamma-ray bursts, would clearly help a lot to check whether the model is well-suited at this stage.
    \item[-] From intermediate time up to our epoch, the model appears as a Chaplygin gas. Particularly, the model acts as a GCG solution and agrees with late time observations. This behaviour can also reconcile the matter-dominated phase with the dark energy-dominated phase, being compatible with our expectations.
\end{itemize}

\subsection{Thermodynamics of matter with pressure}
Matter exhibiting pressure behaves as an effective dark energy contribution. Any evolving dark energy components comply with the laws of thermodynamics and might be described as perfect fluids, particularly in the context of background cosmology. Specifically, the role of specific heats in cosmology has been examined in light of observational data \cite{Luongo:2012dv}, showing that dark energy exhibits strange behavoiur for specific heats, similar to the case of black holes, where the specific heat appears negative \cite{2017CQGra..34g5005B}.

Our EoS mimics dark energy as due to the different thermodynamic stages throughout the universe evolution. This may be prompted easily, emphasising how it can extend the standard $\Lambda$CDM model, by simply working out the adiabatic indices coming from the requirement that heat capacities are constructed in equilibrium thermodynamics for the matter fluid itself, namely assuming they can evolve with time.

As in standard thermodynamics, the heat capacities conform to the relationship between internal energy, enthalpy, and so, in view of the fact that the thermal exchange process is purely adiabatic, implying that the volume scales as $V\propto a^3$, the inferred dark energy contribution exhibits weak interactions with the other constituents, behaving akin to a gaseous fluid source in Einstein's equations.

So, we consider the possibility that the adiabatic index, denoted as $\gamma$, can take specific values while excluding regions where it is not allowed to vary.
To do so, we resort the definition of internal energy and enthalpy,
\begin{subequations}
    \begin{align}
U&=\rho V\,,\\
h&=(\rho+P)V\,,
    \end{align}
\end{subequations}
respectively, which are positive definite functions of the volume $V$, the pressure $P$, and the temperature $T$. Since all state variables evolve with $z$, it is natural to assume the simplest hypothesis where $U$ and $h$ are solely functions of $T$, so that $\rho=\rho(T)$ and $P=P(T)$.
Adopting the standard definition $V=V_0a^{3}$, which captures the evolution of the universe at early and late times, and the relations $H^\prime/H \equiv (1+q)(1+z)^{-1}$ and $H^{\prime\prime}/H \equiv (j-q^2)(1+z)^{-2}$, we define
\begin{subequations}
\begin{align}
C_P&=\frac{2V_0}{T'}\frac{(j-1)}{(1+z)^4}H^2\,,\label{calorecosmografico1}\\
C_V&=\frac{3V_0}{T'}\frac{(2q-1)}{(1+z)^4}H^2\,.\label{calorecosmografico2}
\end{align}
\end{subequations}
where $q$ is the deceleration parameter and $j$ the jerk parameter, related to the so-called cosmography of the universe \cite{Luongo:2011zz,2014PhRvD..90d3531A,2016IJGMM..1330002D,2012PhRvD..86l3516A,2017PDU....17...25A} and the prime here indicates derivative with respect to the redshift, $z$. Then, combining these relations among them, the adiabatic index $\gamma$ yields
\begin{equation}\label{adiabaticindex}
\gamma = \frac{2(j-1) }{3(2q-1)}\,.
\end{equation}
The three allowed regimes are: $0<\gamma<1$, $\gamma=1$ and $\gamma>1$ and so the consequences on the thermodynamics of dark energy are summarised as follows: $C_V,C_P<0$ in the first case, $C_P=0$ and
$C_V<0$ in the second case and $C_V,C_P>0$ in the last case, and furthermore there also exists a region for which\footnote{In general, this could happen at a redshift $z\gg1$, under the hypothesis of de-Sitter contribution to dark energy.} $C_P=0$ and $C_V=0$, occurring as $q\to 1/2$ and $j\to 1$.
Therefore, we have
\begin{equation}
    \gamma = \frac{C_P}{C_V} = \frac{(\rho'+P')V + (\rho + P)V'}{\rho' V + \rho V'}\,,\label{gamma_CP_CV}
\end{equation}
thus, for each epoch the Murnaghan pressure changes accordingly as we list below.

\begin{itemize}
    \item[-] {\bf At primordial time}:
\begin{subequations}\label{condizioni}
        \begin{align}
    q\simeq 1\,,\\
    j\simeq3\,,\\
    \gamma= \frac{4}{3}\,.
        \end{align}
    \end{subequations}
    Here, the adiabatic index is positive-definite, as a consequence of the jerk parameter \cite{Luongo:2013rba}. This indicates an effective polytropic fluid that acts as a cosmological constant plus the contribution of matter and radiation. Clearly, the effective value of the cosmological constant involved in our approximation cannot be exactly that of the $\Lambda$CDM model, since it differs due to the offset imposed from the beginning into the Murnhagam EoS. Further, a slight variation in time is also expected, differently from the standard cosmological model, albeit leaving unaltered the signs of $C_P$ and $C_V$.
    \item[-] {\bf At early times}: the model appears as a logotropic fluid, but since at early times we are at very high redshift, again conditions \eqref{condizioni} are fulfilled. Hence, once again the model resembles the findings of the standard cosmological model even at this stage, implying a suitable thermodynamics as the specific heats are well-defined. Thus, the condition on $\gamma$ suggests that very slight departures occur and, consequently, weak changes at the level of perturbations are also expected.
    \item[-] {\bf At our time}: the model appears as a GCG. Here, the conditions got in Eqs.~\eqref{condizioni} are clearly violated, having
    \begin{subequations}
        \begin{align}
    q\simeq -0.5\,,\\
    j\geq 1\,,\\
    \gamma<0 \,.
        \end{align}
    \end{subequations}
Here the model works very differently than the previous case, indicating a negative adiabatic index. This thermodynamic instability has been explored in view of dark energy and appears related to the form of the two specific heats. Specifically, $C_P>0$ and $C_V<0$, showing that the dark energy fluid behaves very differently than a standard constituent, although the Mayer relation, $C_P-C_V>0$ is fully preserved, showing that dark energy is approximately described by means of a perfect gas.
\end{itemize}
In both the latter cases, the model shows that it corresponds to a  polytropic fluid-like with effective pressure, rescaled by a constant term, with $\gamma\simeq \alpha$. The model is therefore long-ranging, when a specific heat is negative, exhibiting a behaviour that changes significantly as the universe temperature is modified accordingly. The effect of expansion tends to mostly modify the overall form of $P$, changing consequently the physics associated with the fluid itself. The effects of this behaviour are however more evident at intermediate and small redshifts, while at smaller ones the fluid acts as a matter-like component with small pressure. In this respect, it is possible to model the fluid by assuming a double polytrope. Models like this are written as \cite{2016PhRvD..94h3525D}

\begin{equation}
P=P_1\rho^{\gamma_1}+P_2\rho^{\gamma_2}\,,
\end{equation}
with $P_1,P_2,\gamma_1,\gamma_2$ constants to be defined. The Murnagham fluid appears therefore a limiting case that occurs as $\gamma_1<0$ and $\gamma_2=0$, identifying $P_2$ with the constant value of vacuum energy.

For the sake of completeness, there is a further consideration to take into account. The value of vacuum energy, associated to primordial quantum fluctuations, imposed since the very beginning is \emph{fine-tuned} since it cannot be related to the very large values of quantum fields, but rather to the value measured by the Planck satellite \cite{collaboration2020planck}. This is related to the cosmological constant problem, namely the inability to cancel out the degrees of freedom related to high-energy scales associated with quantum fluctuations before and after the phase transition, described in the beginning, at the same time \cite{2012CRPhy..13..566M}. In other words, our model predicts a matter-like component that arose \emph{only after a fine-tuning mechanism of cancellation has occurred to delete the cosmological constant}. Hence, our model \emph{does not} solve the cosmological constant problem by itself, but rather it assumes that it has been resolved somehow during the transition. A possible explanation toward the value of the constant may arise investigating the microphysics of the Murnaghan EoS, i.e., its quantum origin. This is however beyond the purpose of this work and may be subject of further investigations. For additional details about the kind of cancellation mechanisms, expected to erase the cosmological constant, one can refer to Refs.~\cite{2023arXiv230704739B,2023arXiv230801712A,2021JCAP...03..074L,Belfiglio:2023eqi,2022CQGra..39s5014D,2018PhRvD..98j3520L}.

\section{Impact of matter with pressure on linear perturbations}\label{sezione6}

\begin{figure}
\includegraphics[width=\hsize,clip]{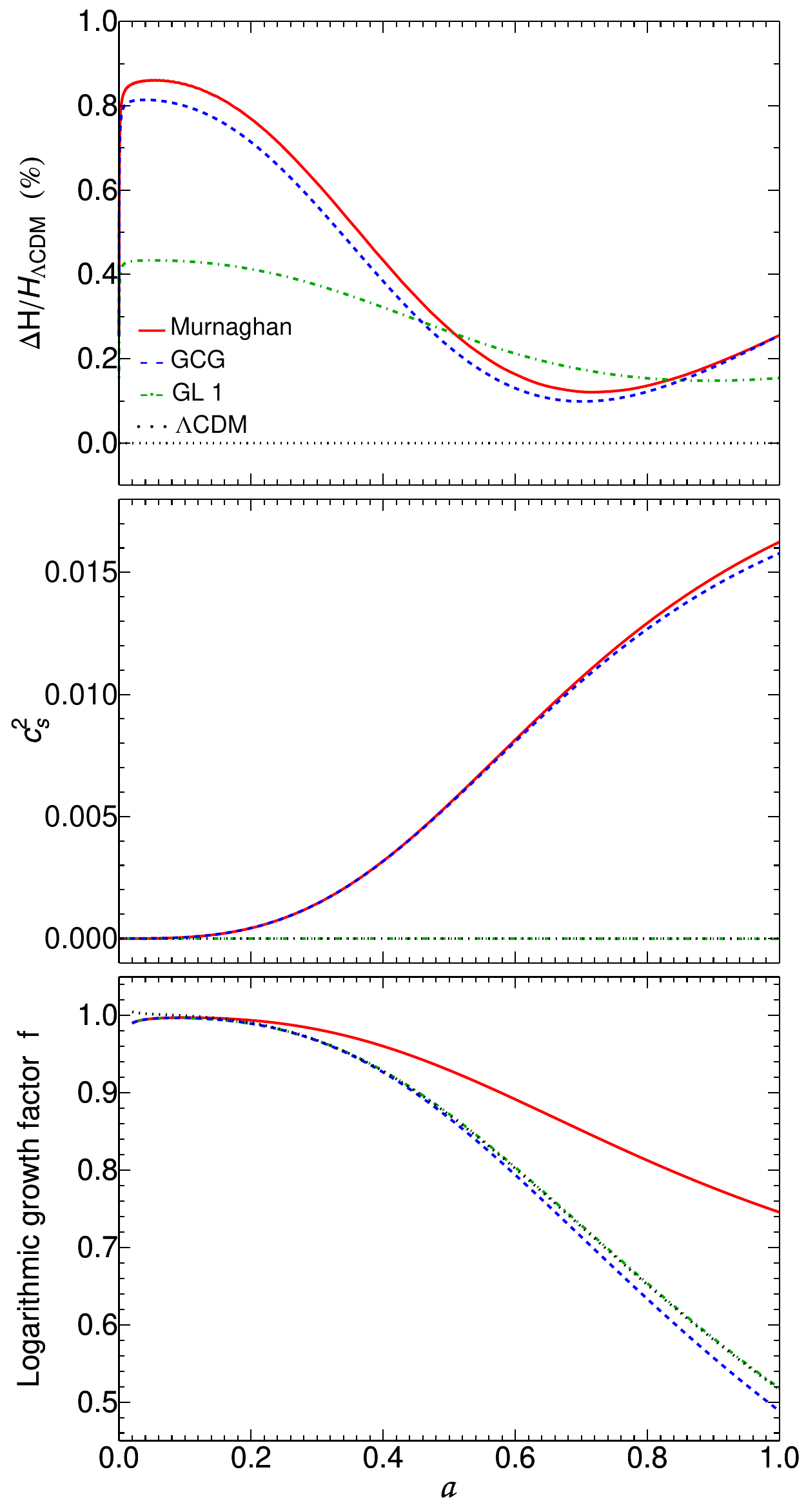}
\caption{Numerical comparison among Murnaghan, GCG, GL1, and $\Lambda$CDM models. \emph{Top panel}: Hubble rate relative differences (with respect to the $\Lambda$CDM one). \emph{Middle panel}: squared sound speeds. \emph{Bottom panel}: logarithmic growth  factors. The parameters used in the plots are taken from Tab.~\ref{tab:results}.}
\label{fig:pert}
\end{figure}

To form structures in the universe requires small initial density perturbations, most likely due to the clustering of matter.
In the linear regime, assuming a homogeneous and isotropic universe \cite{weinberg1972gravitation}, the evolution of such matter density perturbations can be express by the matter fluid $\delta=\delta\rho_{\rm M}/\rho_{\bf M}$ and it is described by \cite{2021PhRvD.104b3520B}
\begin{equation}\label{eq:gf}
 \delta^{\prime\prime} + \left(S\delta\right)^{\prime} +
 \left(2+\frac{H^{\prime}}{H}-\frac{T^{\prime}}{T}\right)
 (\delta^{\prime}+S\delta) - TW\delta = 0\,,
\end{equation}
with $S=3(s-w)$, $T=1+w$, and $W=3(1+3s)\Omega(a)/2$.
The sound speed entering into perturbations is the adiabatic one and its effect is taken into account by
\begin{equation}
    \label{ad_sound_speed}
    s=c_{\rm s}^2=\left(\frac{\partial P}{\partial a} \right) \left(\frac{\partial\rho_{\rm M}}{\partial a}\right)^{-1}\,.
\end{equation}
Defining the matter component density parameter of the perturbed fluid $\Omega(a)$ and the barotropic index $w$,
\begin{subequations}
\begin{align}
    \Omega(a) &= \frac{\rho_{\rm M}(a)}{E(a)^2}\,,\\
    \label{ad_sound_speed}
    w(a) & = -1-\frac{2}{3}\frac{H^\prime(a)}{H(a)}\,,
\end{align}
\end{subequations}
and introducing the logarithmic growth factor $f=(\ln{\delta})^\prime$, Eq.~\eqref{eq:gf} becomes \cite{2021PhRvD.104b3520B}
\begin{equation}
\label{eqn:exact_f}
f^{\prime}+ \left(2 + f + \frac{H^{\prime}}{H}-\frac{T^{\prime}}{T}\right)\left(f+S\right) + S^\prime - TW = 0\,.
\end{equation}
The general approach to the solution of Eq.~\eqref{eqn:exact_f} consists of the steps below \cite{2021PhRvD.104b3520B}.
\begin{itemize}
    \item[1.] The clustering component is assumed to be the dark matter, so one can set $w=0$.
    \item[2.] The phenomenological solution $f \approx \Omega_{\rm M}^{\gamma}(a)$ \citep{Paul}, where $\gamma$ is the so-called growth index,  enables to solve a first-order differential equation for $\gamma$.
    \item[3.] To linearize the above differential equation in $\gamma$, the matter component is assumed to be $\Omega_{\rm m}\approx\mathcal{O}(1)$.
    \item[4.] A vanishing sound speed is assumed through $s=0$.
\end{itemize}
In the following, we take only the above step $1$ and proceed with the numerical evaluations for $\Lambda$CDM, GL1, GCG, and numerical Murnaghan models.
The numerical comparisons of the Hubble rates, the sound speeds, and the logarithmic growth factors of the Murnaghan, GCG, GL1, and $\Lambda$CDM models are portrayed in Fig.~\ref{fig:pert}. The parameter values used in the plots of the $\Lambda$CDM, GL1 and GCG models are taken from Tab.~\ref{tab:results}; the parameters for the Murnaghan model are taken from Eqs~\eqref{consMurnpar}--\eqref{consMurncos}.

The evolution of the Hubble parameter in Fig.~\ref{fig:pert} (top panel) is portrayed as the relative departure with respect to the reference $\Lambda$CDM model. For all models the departure within $1\%$ error for $0\leq a\leq1$.

The most interesting quantity for the evolution of perturbations is the adiabatic sound speed, which is shown in Fig.~\ref{fig:pert} (middle panel).
The sound speed determines the stability and the validity of a given model against perturbations.

The evolution of the growth factor $f$ is depicted in the bottom panel of Fig.~\ref{fig:pert}.

The main findings can be easily listed as follows:

\begin{itemize}
    \item[-] All the models derived from the Murnagham EoS have a roughly negligible and positive-defined sound speeds (identically zero only for the $\Lambda$CDM model).
    \item[-]  At very early times, for all models, $f$ flattens and is of the order unity as expected, then it tends to decrease.
    \item[-]  At late times, the small departures of GL1 and GCG models from the $\Lambda$CDM one are due to the very small (but non-negligible) values of the sound speeds for these two models.
\item[-]  On the contrary, for the Murnaghan EoS we have a large deviation from the $\Lambda$CDM paradigm, resulting in a more efficient growth of perturbations.
This deviation cannot be explained by the sound speed of this model, since it is comparable to the sound speed of the GCG model (see the middle panel of Fig.~\ref{fig:pert}).
The main responsible for this effect is the density parameter of the perturbed fluid $\Omega(a)$. Unlike the analytic case of the GCG model, the Murnaghan EoS does not allow us to disentangle the contribution of the dark matter density from the dark fluid density, therefore, $\Omega(a)$ accounts for the total dark fluid density enhancing the growth of perturbations.
\end{itemize}
\section{Final outlooks and perspectives}\label{sezione7}

We investigated a class of generalised unified dark energy models in which a non-minimal coupling between a tachyonic and scalar fields is introduced. In particular, we demanded that the second field, coupled to the tachyon environment, carries vacuum energy, as predicted by the standard model of particle physics. As a consequence of this recipe, we considered that the second scalar field determines the existence of a vacuum energy term, invoking a corresponding non-removable offset imposed from a symmetry breaking mechanism.

After the transition, in which the second field appears no longer dynamical, the corresponding EoS appears composed as a sum between a GCG and cosmological constant contribution. We reinterpreted this result in view of the so-called Murnagham EoS, widely-adopted in contexts of solid state physics to characterise those fluids, which under the action of external pressure, tend to act against the pressure itself.

We studied the corresponding dynamics of this model and emphasised the main differences with respect to the GCG. Specifically, since the model is not analytically integrable, we argued how it can reduce to particular cases, such as to the $\Lambda$CDM paradigm, to a logotropic fluid and to a Chaplygin gas. To do so, we analysed each approximation and constrained the corresponding bounds over the free parameters in order to to fix them.

Numerical results are taken into account, showing that the model adapts well to cosmic data, as demonstrated by numerical analyses performed by means of MCMC analyses, based on the Metropolis-Hastings algorithm. Using cosmic data, such the OHD, the \textit{Pantheon} SNe Ia and the BAO catalogs, we found limits over the parameters that suggest how the model changes throughout the universe's expansion history.

Precisely, at very early times, we found that the model acts as a logotropic fluid that reduces to the $\Lambda$CDM model, while from early to intermediate times, and up to our time, the model appears to described by a GCG. We interpreted our findings in terms of thermodynamics, showing where the model resembles a polytrope and discussing the role of specific heats.

We the determined the evolution of the density contrast in the linear regime. According to our findings about the possible approximations inferred from the Murnagham EoS, we found that in all cases a non-zero sound speed is accounted for. At very early times, the growth factor flattens and is of the order unity as expected, then it tends to decrease. At late times, the small departures of GL1 and GCG models from the $\Lambda$CDM one are due to the very small (but non-negligible) values of the sound speeds, while for the Murnaghan EoS we found a large deviation from the $\Lambda$CDM paradigm, resulting in a more efficient growth of perturbations. We discussed this deviation in view of current observations and concluded that our model is viable in describing structure formation.

Consequently, we conclude that our framework appears to be well-suited for describing the universe through the action of a single fluid, namely a matter-like fluid with pressure. In other words, we argued that our approach seems to be a suitable candidate to unify dark energy and dark matter under the same standards, applying the non-minimal coupling between two fields, in which at least one of them transports vacuum energy.

Future developments will deal with non-linear perturbations of the fluid itself. We will also focus on how the thermodynamics of this fluid can justify the negative sign of the pressure. Moreover, we will focus on possible generalisations of this recipe in order to obtain a quantum origin of the Murnaghan EoS, capable of explaining the cosmological constant problem and then to obtain more information about the nature of unified dark energy models.

\section*{Acknowledgements}
The work of OL is  partially financed by the Ministry of Education and Science of the Republic of Kazakhstan, Grant: IRN AP19680128. MM acknowledges the support of INFN, iniziativa specifica MoonLIGHT for financial support.

\end{document}